\documentclass[article,nofootinbib,notitlepage]{revtex4-1}
\usepackage{amsmath}
\usepackage[dvips]{graphicx}
\usepackage{amsthm,latexsym,amssymb,amsfonts,array}

\begin{document}

\title{Untangling the Newman-Janis algorithm}
\author{Rafael Ferraro\medskip}
\email{ferraro@iafe.uba.ar}
\thanks{member of Carrera del Investigador Cient\'{\i}fico (CONICET,
Argentina). }

\begin{abstract}
Newman-Janis algorithm for Kerr-Newman geometry is reanalyzed in the
light of Cartan calculus.
\end{abstract}

\affiliation{Instituto de Astronom\'\i a y F\'\i sica del Espacio,
Casilla de Correo 67, Sucursal 28, 1428 Buenos Aires, Argentina}
\affiliation{Departamento de F\'\i sica, Facultad de Ciencias
Exactas y Naturales, Universidad de Buenos Aires, Ciudad
Universitaria, Pabell\'on I, 1428 Buenos Aires, Argentina
\bigskip }

\maketitle

\section{Introduction}
Two years after the discovery of Kerr geometry \cite{Kerr63}, Newman
and Janis showed an algorithm for converting Schwarzschild geometry
into Kerr geometry. They described it as a {\it complex coordinate
transformation on the Schwarzschild metric} for ``deriving'' (quoted
by the authors) the Kerr metric \cite{NJ65}. Of course, the reasons
why such a short cut to Kerr solution does work can be traced to the
behavior of Einstein equations \cite{Schiffer73,Finkelstein75}. On
another hand, Newman-Janis algorithm resembled the complex shift of
the origin used in electromagnetism for obtaining a magnetic dipole
starting from an electric monopole \cite{Newman73}. Actually, Newman
and Janis tried to interpret Kerr geometry by taking advantage of
such analogy. Besides, Newman has shown that the Weyl tensor of
Schwarzschild and Kerr solutions are just different ``real slices''
of a same complex field in complex Minkowski space-time
\cite{Newman73b} (see also Ref.~\cite{Newman88}). More recently,
Newman-Janis algorithm has been invoked to explore axially symmetric
inner solutions \cite{Herrera82,Drake97,Drake00,Viaggiu06} or vacuum
solutions in theories of modified gravity \cite{Yunes13}. Deepening
the understanding of Newman-Janis algorithm can help to improve the
chances of successfully applying this mechanism, or a similar one,
to get axially symmetric solutions in other areas (for instance,
Kaluza-Klein theory, string theory, non-Abelian BH's, alternative
gravities, non-linear electrodynamics, etc.).

The route to Kerr geometry is highly simplified within the framework
of Cartan calculus. We will reobtain the Kerr solution by exploiting
the power of exterior calculus and keeping the focus on the
Newman-Janis mechanism. In Section \ref{sec:II} we introduce a
simple rule to connect two different null tetrads in Minkowski
space-time; these tetrads are based, respectively, on spherical and
twisted spheroidal coordinates. In Section \ref{sec:III} we review
general relativity in terms of null tetrads and the spin connection
as the potentials for torsion and curvature respectively. In Section
\ref{sec:IV} we obtain the Kerr-Newman solution in the Kerr-Schild
form, and state the Newman-Janis algorithm. In Section \ref{sec:V}
we display the conclusions.

\section{Coordinates and tetrads in flat space}\label{sec:II}

\subsection{Twisted spheroidal coordinates}
Let be $x,y,z$ Cartesian coordinates in Euclidean space. The oblate
spheroidal coordinates $r,\theta ,\varphi $ (a case of ellipsoidal
coordinates) are defined as
\begin{equation}
x\ =\ \sqrt{r^{2}+a^{2}}\ \sin \theta \ \cos \varphi \ ,\ \ \ \ \ \ y\ =\
\sqrt{r^{2}+a^{2}}\ \sin \theta \ \sin \varphi \ ,\ \ \ \ \ \ z\ =\ r\ \cos
\theta \ .  \label{oblate}
\end{equation}
The surfaces $r=constant$ and $\theta =constant$ are spheroids and one-sheet
hyperboloids; in fact,
\begin{equation}
\frac{x^{2}+y^{2}}{r^{2}+a^{2}}+\frac{z^{2}}{r^{2}}\ =\ 1\ ,\ \ \ \
\ \ \ \ \ \ \ x^{2}\,+\,y^{2}\,-\,z^{2}\,\tan ^{2}\theta \ =\
a^{2}\,\sin ^{2}\theta \ .\label{surfaces}
\end{equation}
For $\theta $ going to $\pi /2$, the throat radii of the hyperboloid
goes to $a$. The $z=0$ plane is divided into two regions separated
by a circle of radius $a$: i) for $r=0$ it is $x^{2}+y^{2}=a^{2}\
\sin ^{2}\theta <a^{2}$; ii) for $\theta =\pi /2$ it is
$x^{2}+y^{2}=r^{2}+a^{2}>a^{2}$. If $a=0$ the spheroidal coordinates
become spherical; the surfaces $r=constant$ and $\theta =constant$
become spheres and cones.

Oblate spheroidal coordinates (\ref{oblate}) are
Boyer-Lindquist-like coordinates in Euclidean geometry since
\begin{equation}
dx^{2}\,+\,dy^{2}\,+\,dz^{2}\ =\ \frac{r^{2}+a^{2}\ \cos
^{2}\theta}{r^{2}+a^{2}}\ dr^{2}\,+\,(r^{2}+a^{2}\cos ^{2}\theta )\
d\theta ^{2}\,+\,(r^{2}+a^{2})\ \sin ^{2}\theta \ d\varphi ^{2}\ .
\label{distance}
\end{equation}
In this expression, let us replace the azimuth angle $\varphi$ for
\begin{equation}
\phi \ =\ \varphi \,-\,\arctan \,\frac{r}{a}\ \ \ \ \ \Rightarrow \ \ \ \ \
d\phi \ =\ d\varphi \,-\,\frac{a\,dr}{r^{2}+a^{2}}  \label{twisting}
\end{equation}
(then, $\phi $ coincides with $\varphi $ in the region $x^{2}+y^{2}<a^{2}$
where $r=0$). By using the twisted azimuth angle $\phi $, the coefficient of
$dr^{2}$ in the distance (\ref{distance}) becomes equal to $1$:
\begin{equation}
dx^{2}\,+\,dy^{2}\,+\,dz^{2}\ =\ (dr\,+\,a\,\sin ^{2}\theta \ d\phi
)^{2}\,+\,(r^{2}+a^{2}\cos ^{2}\theta )\ (d\theta ^{2}\,+\,\sin ^{2}\theta \
d\phi ^{2})\ .  \label{twisted}
\end{equation}
If $a=0$ one still gets the Euclidean metric in spherical
coordinates. In sum, to change from spherical coordinates to {\it
twisted} spheroidal coordinates in Euclidean space we can follow the
short cut
\begin{eqnarray}
r^{2} &\longrightarrow &\ \rho ^{2}\,\doteq \,r^{2}+a^{2}\cos ^{2}\theta \ ,
\notag \\
dr\ &\longrightarrow &\ dr\,+\,a\,\sin ^{2}\theta \ d\phi \ .
\label{rule1}
\end{eqnarray}

\subsection{Orthonormal and null tetrads}
Tetrads are bases in the cotangent space; they are made up of four
1-forms. Each basis $\{\mathbf{E}^{i}\}$ in the cotangent space is
\textit{dual} of a vector basis $\{\mathbf{E}_{j}\}$ in the tangent
space; this means that $\mathbf{E}^{i}(\mathbf{E}_{j})=\delta
_{j}^{i}$. Both related bases can be expanded in dual coordinate
bases $\{dx^{\mu }\}$, $\{\partial _{\mu }\}$:
\begin{equation}
\mathbf{E}^{i}\,=\,E_{\mu }^{i}\ dx^{\mu }\
,\hspace{3cm}\mathbf{E}_{i}\,=\,E_{i}^{\mu }\ \partial _{\mu }\ .
\end{equation}%
Duality implies that the matrix $E_{i}^{\mu }$ is inverse of $E_{\mu }^{i}$.

The metric properties of a manifold can be represented by the metric
tensor field $\mathbf{g}$ or, alternatively, a field of tetrads
$\{\mathbf{e}^{\hat{a}}\}$ linked to the metric by the assumption of
orthonormality:
\begin{equation}
g_{\mu \nu }\ =\ \eta _{\hat{a}\hat{b}}\ e_{\mu }^{\hat{a}}\ e_{\nu
}^{\hat{b}}\ ,\hspace{3cm}g^{\mu \nu }\ =\ \eta ^{\hat{a}\hat{b}}\
e_{\hat{a}}^{\mu }\ e_{\hat{b}}^{\nu }\ ,  \label{metric}
\end{equation}%
where $\eta _{\hat{a}\hat{b}}$ is the Minkowskian metric
$\text{diag}(1,-1,-1,-1)$. We can check the orthonormality by
computing the inner products between elements of the basis
$\{\mathbf{e}^{\hat{a}}\}$:
\begin{equation}
\mathbf{e}^{\hat{a}}\cdot \mathbf{e}^{\hat{b}}\ =\ g^{\mu \nu
}\,e_{\mu }^{\hat{a}}\,e_{\nu }^{\hat{b}}\ =\ \eta
^{\hat{a}\hat{b}}\ ,
\end{equation}
because of duality. We avoid coordinate indexes by writing the
relation (\ref{metric}) as
\begin{equation}
\mathbf{g\;}=\;\eta _{\hat{a}\hat{b}}\ \mathbf{e}^{\hat{a}}\otimes
\mathbf{e}^{\hat{b}}\ ,  \label{metrica}
\end{equation}
where $\otimes$ stands for the tensor product.

At each point of the manifold there exists a continuous of
orthonormal tetrad fields, all of them related through (local)
Lorentz transformations $L$,
\begin{equation}
\mathbf{e}_{\hat{a}^{\prime }}\ =\ L_{\ \hat{a}^{\prime }}^{\hat{a}}\
\mathbf{e}_{\hat{a}}\ ,\hspace{3cm}\mathbf{e}^{\hat{a}^{\prime }}\ =\ L_{\
\hat{a}}^{\hat{a}^{\prime }}\ \mathbf{e}^{\hat{a}}\,  \label{lorentz}
\end{equation}
(duality requires that $L_{\ \hat{a}}^{\hat{a}^{\prime }}$ be
inverse of $L_{\ \hat{a}^{\prime }}^{\hat{a}}$). This kind of
ambiguity has no consequences for the metric $\mathbf{g}$ because
$\eta _{\hat{a}\hat{b}}$ is Lorentz-invariant.

The link between metric and tetrad can also be established through a
null tetrad. Both strategies are related, since any orthonormal
tetrad $\{\mathbf{e}^{\hat{a}}\}$ defines a null tetrad
$\{\mathbf{n}^{a}\}=\{\mathbf{l },\ \mathbf{n},\ \mathbf{m},\
\overline{\mathbf{m}}\}$:
\begin{equation}
\mathbf{l}\,=\,\frac{1}{\sqrt{2}}\
(\mathbf{e}^{\hat{0}}+\mathbf{e}^{\hat{1} })\
,\;\;\;\mathbf{n}\,=\,\frac{1}{\sqrt{2}}\
(\mathbf{e}^{\hat{0}}-\mathbf{e} ^{\hat{1}})\
,\;\;\;\mathbf{m}\,=\,\frac{1}{\sqrt{2}}\ (\mathbf{e}^{\hat{2}
}+i\;\mathbf{e}^{\hat{3}})\
,\;\;\;\overline{\mathbf{m}}\,=\,\frac{1}{\sqrt{2}} \
(\mathbf{e}^{\hat{2}}-i\;\mathbf{e}^{\hat{3}})\ .\;  \label{lnmm}
\end{equation}
In fact, the tetrad (\ref{lnmm}) results to be null:
\begin{equation}
\mathbf{l}\cdot \mathbf{l}\,=0\,=\mathbf{n}\cdot \mathbf{n}\
,\;\;\;\;\;\;\;\;\; \mathbf{m}\cdot
\mathbf{m}\,=\,0\,=\,\overline{\mathbf{m}}\cdot \overline{
\mathbf{m}}\ ;
\end{equation}
besides it is
\begin{equation}
\mathbf{l}\cdot \mathbf{n}\,=\,1\,=\,-\mathbf{m}\cdot
\overline{\mathbf{m}}\ ,\;\;\;\;\;\;\;\;\;\mathbf{l}\cdot
\mathbf{m}\,=\,0\,=\,\mathbf{n}\cdot \mathbf{m} \ .
\end{equation}
The relationship between related orthonormal and null tetrads can be written
as
\begin{equation}
\mathbf{n}^{a}\ =\ \Lambda _{\ \hat{a}}^{a}\ \mathbf{e}^{\hat{a}}\,\
,\;\;\;\;\;\;\;\;\;\mathbf{e}^{\hat{a}}\ =\ \Lambda _{\
a}^{\hat{a}}\ \mathbf{n}^{a}\,\ ,
\end{equation}
where $\Lambda _{\ \ \hat{a}}^{a}$ and its inverse matrix are
\begin{equation}
\Lambda _{\ \hat{a}}^{a}\ =\ \frac{1}{\sqrt{2}}\ \left(
\begin{array}{rrrr}
1 & 1 & \ \ 0 & 0 \\
1 & -1 & \ \ 0 & 0 \\
0 & 0 & \ \ 1 & i \\
0 & 0 & \ \ 1 & -i
\end{array}%
\right) \;,\;\;\;\;\;\;\;\;\;\Lambda _{\ a}^{\hat{a}}\ =\
\frac{1}{\sqrt{2}}\ \left(
\begin{array}{rrrr}
1 & 1 & 0 & \ \ 0 \\
1 & -1 & 0 & \ \ 0 \\
0 & 0 & 1 & \ \ 1 \\
0 & 0 & -i & \ \ i
\end{array}
\right) \;.  \label{matrices}
\end{equation}%
By replacing in Eq.~(\ref{metrica}) one gets
\begin{equation}
\mathbf{g}\ =\ \eta _{\hat{a}\hat{b}}\ \Lambda _{\ a}^{\hat{a}}\
\Lambda _{\ b}^{\hat{b}}\ \mathbf{n}^{a}\otimes \mathbf{n}^{b}\ =\
\eta _{ab}\,\ \mathbf{n}^{a}\otimes \mathbf{n}^{b}\,,
\end{equation}%
where%
\begin{equation}
\eta _{ab}\ =\ \eta _{\hat{a}\hat{b}}\ \Lambda _{\ a}^{\hat{a}}\ \Lambda _{\
b}^{\hat{b}}\ =\ \left(
\begin{array}{rrrr}
0 & \ \ 1 & 0 & 0 \\
1 & \ \ 0 & 0 & 0 \\
0 & \ \ 0 & 0 & -1 \\
0 & \ \ 0 & -1 & 0
\end{array}
\right) \ =\ \,\eta ^{ab}\ .
\end{equation}
In sum, the relation between metric tensor and null tetrad is
\begin{equation}
\mathbf{g}\ =\ \mathbf{l}\otimes \mathbf{n}\,+\,\mathbf{n}\otimes
\mathbf{l}\,-\, \mathbf{m}\otimes
\overline{\mathbf{m}}\,-\,\overline{\mathbf{m}}\otimes \mathbf{m}\,.
\label{metricalnmm}
\end{equation}
For instance, according to Eq.~(\ref{twisted}), a possible null
tetrad for Minkowski space-time is
\begin{eqnarray}
&&\mathbf{l}\ =\ \frac{1}{\sqrt{2}}\ ( dt+dr\,+\,a\,\sin ^{2}\theta
\ d\phi )\, ,\;\;\;\;\;\;\mathbf{n}\ =\ \frac{1}{\sqrt{2}}\ (
dt-dr\,-\,a\,\sin ^{2}\theta \ d\phi ) \, ,  \notag
\\ \label{tetradanula}
\\ \notag
&&\mathbf{m}\ =\ \frac{\xi}{\sqrt{2}}\ (d\theta \,+\,i\,\sin \theta
\ d\phi )\,
,\;\;\;\;\;\;\;\;\;\;\;\;\;\;\;\;\;\;\overline{\mathbf{m}}\ =\
\frac{\overline{\xi}}{\sqrt{2}}\,\ (d\theta \,-\,i\,\sin \theta \
d\phi )\, ,
\end{eqnarray}
where $r$, $\theta$, $\phi$ are twisted spheroidal coordinates and
$\xi \doteq r\,+\,i\ a\,\cos \theta $. Thus, the rule (\ref{rule1})
for passing from spherical ($a=0$) to twisted spheroidal coordinates
is rephrased as \medskip
\begin{eqnarray}
&&\text{i) in the }\{\mathbf{m,\overline{\mathbf{m}}}\}\text{
sector, replace \ \ \ \ \ \ \ \ \ \ \ \ }r\ \longrightarrow \
\xi\,,\
\overline{\xi} \notag \\
\label{rule2}
\\
&&\text{ii) in the }\{\mathbf{l,n}\}\text{ sector, replace \ \ \ \ \
\ \ \ \ \ \ \ \ }dr\ \longrightarrow \ dr\,+\,a\,\sin ^{2}\theta \
d\phi \,\,. \notag
\end{eqnarray}
\vskip.3cm\parindent=0pt Eq.~(\ref{twisted}) shows that
$dr+a\,\sin^2\!\theta\,d\phi$ is a unitary 1-form on each $t=${\it
const.} hypersurface, being orthogonal to $\mathbf{m}$ and
$\overline{\mathbf{m}}$. Actually it is the covector associated with
the unitary vector $\partial/\partial r$ tangent to the lines
$\theta , \phi =${\it const.}\footnote{It can be verified that lines
$\theta , \phi =${\it const.} are the {\it straight} lines
generating the one-sheet hyperboloids of Eq.~(\ref{surfaces}). They
form a congruence of (geodesic) straight lines displaying the axial
symmetry we will pursue for the gravitational field in Section
\ref{sec:IV}.}

\section{Gravity in Cartan language}\label{sec:III}
Newman-Janis algorithm is not merely a way to change coordinates in
flat space-time. It involves gravity; it is a short cut to change
from Schwarzschild (or Reissner-Nordstrom) geometry to Kerr (or
Kerr-Newman) geometry. So, we should add gravity (curvature) to the
rule (\ref{rule2}). Gravity can be added in Eq.~(\ref{metricalnmm})
by using tetrads other than the one of Eq.~(\ref{tetradanula}).
However, not any tetrad is allowed since the so built new metric
should still accomplish Einstein equations.

\subsection{Torsion and curvature}
Let be $\Gamma_{jk}^i$ the affine connection defining the covariant
derivative in a manifold. Since the derivative index $k$ behaves
tensorially under changes of basis, the $\Gamma_{jk}^i$'s define a
set of 1-forms $\mathbf{\omega }_{\ j}^i$,
\begin{equation}
\mathbf{\omega }_{\ j}^{i}\ \doteq \ \Gamma _{jk}^{i}\ \mathbf{E}^{k}\ ,
\end{equation}
called the spin connection. The transformation of the spin connection under
change of basis,
\begin{equation}
\mathbf{E}^{i^{\prime }}\ =\ \Lambda _{\ \ i}^{i^{\prime }}\
\mathbf{E}^{i}\ \ ,\hspace{1.5cm}\mathbf{\omega }_{\ j^{\prime
}}^{i^{\prime }}\ =\ \Lambda _{\ i}^{i^{\prime }}\ \mathbf{\omega
}_{\ j}^{i}\ \Lambda_{\ j^{\prime }}^{j}\ +\ \Lambda _{\
k}^{i^{\prime }}\ d\Lambda_{\ j^{\prime }}^{k}\ \ ,
\label{transformation}
\end{equation}
allows the preservation of the tensorial character of an object
under covariant differentiation. By differentiating the tetrad and
the connection, one defines two tensor-valued 2-form fields on the
manifold. Torsion $\mathbf{T}^{i}$ is the covariant derivative of
the tetrad,
\begin{equation}
\mathbf{T}^{i}\ \doteq \ D\mathbf{E}^{i}\ = \ d\mathbf{E}^{i}\ +\
\mathbf{\omega}_{\ j}^{i}\wedge \mathbf{E}^{j}\ ,  \label{torsion}
\end{equation}
and curvature $\mathbf{R}_{ \ j}^{i}$ is built with derivatives of
the spin connection:
\begin{equation}
\mathbf{R}_{\ j}^{i}\ \doteq \ d\mathbf{\omega }_{\ j}^{i}\ +\
\mathbf{\omega}_{\ k}^{i}\wedge \mathbf{\omega }_{\ j}^{k}\ .
\label{curvature}
\end{equation}
$\mathbf{T}^{i}$ and $\mathbf{R}_{\ j}^{i}$ have a mixed character. On one
hand they are 2-forms for each choice of their indexes. On the other hand
they transform as components of tensors in the indexes $i,\ j,...$ by virtue
of the behaviors (\ref{transformation}).

In general, the covariant derivative $D$ of a tensor-valued $p$-form
is a $(p+1)$-form that preserves its tensorial character thanks to
the compensating terms contributed by the connection. Tensor
$\mathbf{R}_{\ j}^{i}$ cannot be thought as the covariant derivative
of $\mathbf{\omega }_{\ j}^{i}$ because the connection does not
transform as a tensor. Tensor $\mathbf{R}_{\ j}^{i}$ can be
covariantly differentiated to obtain the (second) Bianchi identity,
\begin{equation}
D\mathbf{R}_{\ j}^{i}\ =\ d\mathbf{R}_{\ j}^{i}+\mathbf{\omega }_{\
k}^{i}\wedge \mathbf{R}_{\ j}^{k}-\mathbf{\omega }_{\ j}^{k}\wedge
\mathbf{R}_{\ k}^{i}\ \equiv \ 0\ .
\end{equation}
For more details about Cartan calculus see, for instance,
Ref.~\cite{Gasperini13}.

\subsection{Einstein equations}
In Gravity we choose an orthonormal tetrad
$\{\mathbf{e}^{\hat{a}}\}$ and the spin connection $\{\mathbf{\omega
}_{\ \ \hat{b}}^{\hat{a}}\}$ to play the role of potentials
describing the gravitational fields (torsion and curvature). The
assumed orthonormality of the tetrad establishes the link
tetrad-metric; this link is invariant under local Lorentz
transformations (\ref{lorentz}). On the other hand the spin
connection is assumed to be \textit{metric}, which means the
vanishing of the covariant derivative of the (Lorentz) tensor-valued
0-form $\eta _{\hat{a}\hat{b}}$:
\begin{equation}
0\ =\ D\eta _{\hat{a}\hat{b}}\ =\ d\eta _{\hat{a}\hat{b}}\ -\ \mathbf{\omega
}_{\ \ \hat{a}}^{\hat{c}}\ \eta _{\hat{c}\hat{b}}\ -\ \mathbf{\omega }_{\ \
\hat{b}}^{\hat{c}}\ \eta _{\hat{a}\hat{c}}\ ,
\end{equation}
i.e.,
\begin{equation}
\mathbf{\omega }_{\hat{b}\hat{a}}\ =\ -\mathbf{\omega }_{\hat{a}\hat{b}}
\end{equation}
(Lorentz tensor indexes are lowered with $\eta _{\hat{a}\hat{b}}$).
This property also implies
\begin{equation}
D\epsilon _{\hat{a}\hat{b}\hat{c}\hat{d}}\ =\ 0\ ,\label{metricity}
\end{equation}
where $\epsilon _{\hat{a}\hat{b}\hat{c}\hat{d}}$ is the Levi-Civita
symbol, which is a tensor under Lorentz
transformations.\footnote{For theories harboring a non-metricity
field see, for instance, Ref.~\cite{Obukhov06}.}

General Relativity is a theory of gravity governed by the
Einstein-Hilbert Lagrangian, which is the Lorentz scalar-valued
4-form (volume) defined as
\begin{equation}
L\ \ =\ \frac{1}{32\,\pi \,G}\ \epsilon
_{\hat{a}\hat{b}\hat{c}\hat{d}}\ \mathbf{e}^{\hat{a}}\wedge
\mathbf{e}^{\hat{b}}\wedge \mathbf{R}^{\hat{c} \hat{d}}\ .
\end{equation}
In Palatini approach, the action is varied independently with
respect to the tetrad and the connection:
\begin{equation}
\delta L\ \ \propto \ \ \epsilon _{\hat{a}\hat{b}\hat{c}\hat{d}}\
\left( 2\ \delta \mathbf{e}^{\hat{a}}\wedge
\mathbf{e}^{\hat{b}}\wedge \mathbf{R}^{\hat{c}\hat{d}}\ +\
\mathbf{e}^{\hat{a}}\wedge \mathbf{e}^{\hat{b}}\wedge D\delta
\mathbf{\omega }^{\hat{c}\hat{d}}\right)
\end{equation}
(remarkably, the difference between two connections does transform
as a tensor). We integrate by parts the second term to get two
(vacuum) dynamical equations,
\begin{equation}
\epsilon _{\hat{a}\hat{b}\hat{c}\hat{d}}\ \mathbf{e}^{\hat{a}}\wedge
\mathbf{T}^{\hat{b}}\ =\ 0\ ,  \label{eqtorsion}
\end{equation}
\begin{equation}
\epsilon _{\hat{a}\hat{b}\hat{c}\hat{d}}\ \mathbf{e}^{\hat{b}}\wedge
\mathbf{R}^{\hat{c}\hat{d}}\ =\ 0\ .  \label{eqeinstein}
\end{equation}
Equations (\ref{eqtorsion}) imply the vanishing of torsion (they are
as many independent equations as independent components of the
torsion). So the connection becomes the Levi-Civita connection,
which is the (antisymmetric) metric connection that cancels out the
torsion. Equations (\ref{eqeinstein}) are Einstein equations.

Einstein equations keep their form when null tetrads are used: since $\det
(\Lambda _{\ \ \hat{a}}^{a})=i$, it is
\begin{equation}
\epsilon _{\hat{a}\hat{b}\hat{c}\hat{d}}\ =\ -i\ \epsilon _{abcd}\
\Lambda _{\ \ \hat{a}}^a\ \Lambda _{\ \ \hat{b}}^b\ \Lambda _{\ \
\hat{c}}^c\ \Lambda _{\ \ \hat{d}}^d\ ;
\end{equation}
besides,
\begin{equation}
\eta ^{\hat{d}\hat{e}}\ \Lambda _{\ \ \hat{d}}^{d}\ \ =\ \ \eta ^{de}\
\Lambda _{\ \ e}^{\hat{e}}\ .
\end{equation}
Thus, vacuum Einstein equations are as well
\begin{equation}
\epsilon _{abcd}\ \mathbf{n}^{b}\wedge \mathbf{R}^{cd}\ \ =\ \ 0\ .
\label{eqeinsteinnull}
\end{equation}

\section{From flat space-time to Kerr geometry}\label{sec:IV}
Newman-Janis algorithm connects Schwarzschild and Kerr geometries by means
of the rules (\ref{rule2}) plus a rule affecting the Newtonian gravitational
potential in Schwarzschild geometry:
\begin{equation}
\frac{2\ M}{r}\ \longrightarrow \ \frac{2\ M\ r}{\rho ^{2}}\ =\ M\left(
\frac{1}{\xi }+\frac{1}{\bar{\xi}}\right) \,.  \label{rule}
\end{equation}
To trace the reasons for this rule, we will introduce gravity by performing
just a tiny change of the Minkowskian tetrad (\ref{tetradanula}); we will
only change the 1-form $\mathbf{l}$:
\begin{equation}
\mathbf{l}\ \longrightarrow \ \mathbf{l}\,\ +\ f(r,\theta )\ \mathbf{n}\,.
\label{change}
\end{equation}
Thus, the new geometry is expressed in the Kerr-Schild form: ${\bf
g} = {\overline{\bf g}} + 2\, f(r, \theta)\, {\bf n}\otimes {\bf
n}$, where ${\overline{\bf g}}$ is the Minkowskian seed metric
(\ref{metricalnmm}, \ref{tetradanula})
\cite{Kerr63,Kerr65,Debney67}. Since the function $f$ depends just
on $(r,\theta )$, the new geometry will remain stationary and
axially symmetric. Coordinates $\{u\doteq t-r,\,r,\,\theta ,\,\phi
\}$ are outgoing Eddington-Finkelstein-like coordinates. While ${\bf
l}$ and ${\bf n}$ are null directions on an equal footing in Eq.
(\ref{tetradanula}) --they are covectors of $\partial/\partial t \pm
\partial/\partial r$, so they represent rays of light traveling in
opposite directions in Minkowski space-time (see Footnote 1)--,
instead they will not be equivalent in the new geometry ${\bf g}$
because gravity will distinguish ingoing and outgoing rays by means
of the function $f(r,\theta )$.

Function $f$ can be regarded as well as $f(\xi ,\overline{\xi})$; it
cannot be arbitrarily chosen because the new geometry must verify
Einstein equations (\ref{eqeinsteinnull}) and preserve the vanishing
of torsion. Function $f(\xi, {\overline\xi})$ will play the role of
gravitational potential of the new geometry, because
$g_{tt}\rightarrow 1+ f(\xi, {\overline\xi})$. Newman-Janis rule
(\ref{rule}) says that both Schwarzschild and Kerr geometries are
written with the same function $f(\xi, {\overline\xi})$; the sole
difference between them comes from the vanishing or not of $a$ in
$\xi = r + i\, a\, \cos\theta$ (apart from the explicit dependence
on $a$ of the original null forms ${\bf l}$ and ${\bf n}$).
Therefore, the good working of Newman-Janis rule requires that
$f(\xi, {\overline\xi})$ fulfills equations in the variables $\xi,
{\overline\xi}$ that do not explicitly contain the parameter $a$. In
fact this will be the case, as we are going to show.

\subsection{Keeping the torsion null}
According to the definition (\ref{torsion}), the vanishing of
torsion is expressed by the equation
\begin{equation}
\ d\mathbf{n}^{a}\ =\ -\mathbf{\omega }^{ab}\wedge \mathbf{n}_{b}\ ,
\label{torsionnull}
\end{equation}
where indexes are lowered and raised with the metric $\eta _{ab}$
and its inverse $\eta ^{ab}$. So, the index $0$ goes to $1,$ and $2$
changes to $3$ plus a change of sign. In particular,
$\{\mathbf{n}_{a}\}=\{\mathbf{n},\ \mathbf{l},\
-\overline{\mathbf{m}},\ -\mathbf{m}\}$. Like
$\mathbf{\omega}_{\hat{a}\hat{b}}$, $\mathbf{\omega }^{ab}$ is
antisymmetric too. This is because the transformation
(\ref{matrices}) is constant;\ then the relation between
$\mathbf{\omega }_{\hat{a}\hat{b}}$ and $\mathbf{\omega }_{ab}$
looks tensorial in Eq. (\ref{transformation}). The antisymmetry of
$\mathbf{\omega }^{ab}$ allows to solve Eq.~(\ref{torsionnull}) for
the components of the torsionless spin connection, so obtaining the
Levi-Civita connection:
\begin{equation}
(\mathbf{\omega }^{ab})^{c}=\frac{1}{2}\ \left[
(d\mathbf{n}^{a})^{bc}+(d\mathbf{n}^{b})^{ca}-(d\mathbf{n}^{c})^{ab}\right]
\ .  \label{LC}
\end{equation}
Since the null tetrad (\ref{tetradanula}) satisfies%
\begin{eqnarray}
d\mathbf{l\ } &=&\mathbf{\ }-d\mathbf{n\ }=\mathbf{\
}-\frac{1}{\sqrt{2}}\ \left( \frac{1}{\xi }-\frac{1}{\overline{\xi
}}\right) \ \mathbf{m}\wedge \overline{\mathbf{m}},  \notag
\\ \label{derivativesoftetrad}
\\
d\mathbf{m\ } &=&\mathbf{\ }\frac{1}{\sqrt{2}\ \xi }\
(\mathbf{l-n)\wedge m\ -\ }\frac{1}{\sqrt{2}\ \overline{\xi }}\left(
\cot \theta -\frac{2i}{\xi }\ a\ \sin \theta \right) \
\mathbf{m}\wedge \overline{\mathbf{m}}\ , \notag
\end{eqnarray}
then by replacing these 2-forms in Eq.~(\ref{LC}) we get the
connection for the Minkowskian basis (\ref{tetradanula}):
\begin{equation}
\ \,\,\mathbf{\omega }^{ab}\ \,=\mathbf{\ }\ \left(
\begin{array}{cccc}
0 & 0 & \frac{\mathbf{m}}{\sqrt{2}\ \overline{\xi }} &
\frac{\overline{\mathbf{m}}}{\sqrt{2}\ \xi } \\
\begin{array}{c}
\\
...
\end{array}
&
\begin{array}{c}
\\
0
\end{array}
&
\begin{array}{c}
\\
\;-\frac{\mathbf{m}}{\sqrt{2}\ \overline{\xi }}\;
\end{array}
&
\begin{array}{c}
\\
-\frac{\overline{\mathbf{m}}}{\sqrt{2}\ \xi }
\end{array}
\\
\begin{array}{c}
\\
...
\end{array}
&
\begin{array}{c}
\\
...
\end{array}
&
\begin{array}{c}
\\
0
\end{array}
&
\begin{array}{c}
\\
\;d\left[ \ln \frac{\xi }{\overline{\xi }}\right] -\frac{\cot \theta
}{\sqrt{2}}\left( \frac{\mathbf{m}}{\xi
}-\frac{\overline{\mathbf{m}}}{\overline{\xi }}\right) \;
\end{array}
\\
\begin{array}{c}
\\
...
\end{array}
&
\begin{array}{c}
\\
...
\end{array}
&
\begin{array}{c}
\\
...
\end{array}
&
\begin{array}{c}
\\
0
\end{array}
\end{array}
\right) \ .  \label{omega}
\end{equation}
As can be seen, there does not exist a short cut to obtain the
result (\ref{omega}) from the Minkowskian spin connection in ($a=0$)
spherical coordinates, since no trace of $\ln (\xi/\overline{\xi})$
remains in the $a=0$ case.

If the geometry is modified by a change $\delta \mathbf{n}^{a}$ of
the null tetrad, then a change of the spin connection must happen as
well to preserve the vanishing of torsion. The new spin connection
$\mathbf{\omega }^{ab}+\delta \mathbf{\omega }^{ab}$ could be
computed by using again the Eq.~(\ref{LC}). However, the issue could
also be considered at the level of Eq.~(\ref{torsionnull}), which
implies a relation between $\delta \mathbf{n}^{a}$ and $\delta
\mathbf{\omega }^{ab}$ in order to keep the torsion null:
\begin{equation}
d\ \delta \mathbf{n}^{a}\ =\ -\, \mathbf{\omega }^{ab}\wedge \delta
\mathbf{n}_{b}\ -\ \delta \mathbf{\omega }^{ab}\wedge
\mathbf{n}_{b}\ -\ \delta \mathbf{\omega }^{ab}\wedge \delta
\mathbf{n}_{b}\ . \label{change1}
\end{equation}%
As said, the change of tetrad we are going to introduce is $\delta
\mathbf{l} =f\ \mathbf{n}$, $\delta \mathbf{n}^{\alpha }=0$ ($\alpha
\neq 0$). We will argue that $\delta \mathbf{\omega }^{ab}$ should
be linear in $f$; besides, as a solution of Einstein equations, $f$
should be proportional to some integration constant measuring the
strength of the gravitational field. Since Eq.~(\ref{change1}) has
to be satisfied for any value of the integration constant, then the
quadratic term must separately cancel out:
\begin{equation}
\delta \mathbf{\omega }^{a1}\wedge \mathbf{n\ }=\mathbf{\ }0\ .\
\label{change4}
\end{equation}
Eq.~(\ref{change1}) for $a=0$ then becomes
\begin{equation}
df\wedge \mathbf{n}+f\ d\mathbf{n}\ \mathbf{\ }=\mathbf{\ -}\ \delta
\mathbf{ \omega }^{0\alpha }\wedge \mathbf{n}_{\alpha }\ ,
\label{change2}
\end{equation}
where we have used that $\mathbf{\omega }^{01}=0$. Eq.
(\ref{change1}) for $ a=\alpha $ is
\begin{equation}
f\ \mathbf{\omega }^{\alpha 1}\wedge \mathbf{n}\ =\ \delta
\mathbf{\omega }^{b\alpha }\wedge \mathbf{n}_{b}  \label{change3}
\end{equation}%
(notice that $\mathbf{n}_{1}=\mathbf{l}$ is the 1-form belonging to the
original tetrad, since the change $\delta \mathbf{l}$ is separately
written). The unknowns $\delta \mathbf{\omega }^{b\alpha }$ must fulfill
\begin{equation}
\delta \mathbf{\omega }^{02}\ =\ \overline{\delta
\mathbf{\omega}^{03}},\;\;\;\;\;\delta \mathbf{\omega }^{12}\ =\
\overline{\delta \mathbf{\omega }^{13}},\;\;\;\;\;\delta
\mathbf{\omega }^{23}\ =\ \overline{\delta \mathbf{\omega }^{32}}\
=\ -\overline{\delta \mathbf{\omega }^{23}},
\end{equation}
as it results from the complex behavior of the null tetrad. One can
start by using Eq.~(\ref{change4}) to eliminate a term in the Eq.
(\ref{change3}) for $\alpha =1$; it results
\begin{equation}
0\mathbf{\ \ }=\mathbf{\ }\ \delta \mathbf{\omega }^{21}\wedge
\overline{\mathbf{m}}\ +\ \delta \mathbf{\omega }^{31}\wedge
\mathbf{m}\ .
\end{equation}
We will try the solution $\delta \mathbf{\omega }^{21}=0=\delta
\mathbf{\omega}^{31}$: by replacing it in Eq.~(\ref{change3}) with
$\alpha =2$, we solve $\delta \mathbf{\omega }^{32}$. Besides, $df$
in Eq. (\ref{change2}) is
\begin{equation}
df\ \mathbf{\ }=\mathbf{\ }\ \partial _{\xi }f\ d\xi \ +\
\partial_{\overline{\xi }}f\ d\overline{\xi }\ ,
\end{equation}
where
\begin{equation}
d\xi \ =\ dr-i\ a\ \sin \theta \ d\theta \ =\
\frac{\mathbf{l}-\mathbf{n}}{\sqrt{2}}\ -\ i\ \sqrt{2}\ a\ \sin
\theta \ \frac{\overline{\mathbf{m}}}{\overline{\xi }}\ .
\label{dxi}
\end{equation}
It is easy to verify that the solution $\delta \mathbf{\omega
}^{ab}$ to Eqs. (\ref{change2}), (\ref{change3}) is
\begin{equation}
\,\delta \mathbf{\omega }^{ab}\ \,=\ \left(
\begin{array}{cccc}
0 & \;\frac{1}{\sqrt{2}}\left( \partial _{\xi }f+\partial
_{\overline{\xi}}f\right) \mathbf{n}\ \  & \ \ \frac{f\
\mathbf{m}}{\sqrt{2}\ \xi }+\frac{i \sqrt{2}\,a\,\sin \theta
}{\overline{\xi }}\,\partial _{\xi }f\,\mathbf{n}\ \  & \ \ \frac{f\
\overline{\mathbf{m}}}{\sqrt{2}\ \overline{\xi }}-\frac{i
\sqrt{2}\,a\,\sin \theta }{\xi }\,\partial _{\overline{\xi
}}f\,\mathbf{n}\;
\\
\begin{array}{c}
\\
...
\end{array}
&
\begin{array}{c}
\\
0
\end{array}
&
\begin{array}{c}
\\
0
\end{array}
&
\begin{array}{c}
\\
0
\end{array}
\\
\begin{array}{c}
\\
...
\end{array}
&
\begin{array}{c}
\\
...
\end{array}
&
\begin{array}{c}
\\
\ 0
\end{array}
&
\begin{array}{c}
\\
\frac{f}{\sqrt{2}}\ \left( \frac{1}{\xi }-\frac{1}{\overline{\xi
}}\right) \ \mathbf{n}
\end{array}
\\
\begin{array}{c}
\\
...
\end{array}
&
\begin{array}{c}
\\
...
\end{array}
&
\begin{array}{c}
\\
...
\end{array}
&
\begin{array}{c}
\\
\ 0%
\end{array}
\end{array}
\right)  \label{deltaomega}
\end{equation}
and Eq.~(\ref{change4}) is accomplished too. Although $\delta
\mathbf{\omega }^{ab}$ explicitly depends on the parameter $a$, we
are going to show that the equations $f(\xi ,\overline{\xi})$
accomplishes do not contain $a$ in an explicit way.

\subsection{Newman-Janis rules}
Function $f(\xi ,\overline{\xi})$, which describes the gravitational
field of the solution under consideration, is dictated by Einstein
equations. The modified geometry, characterized by
$\mathbf{n}^{b}+\delta \mathbf{n}^{b}$,
$\mathbf{\omega}^{ab}+\delta\mathbf{\omega}^{ab}$, has to fulfill
Einstein equations (\ref{eqeinsteinnull}):
\begin{equation}
\epsilon _{abcd}\ (\mathbf{n}^{b}+\delta \mathbf{n}^{b})\wedge
\mathbf{R}^{cd}(\mathbf{\omega} + \delta \mathbf{\omega )}\  = \ 0\
. \label{einsteinchange}
\end{equation}
By expanding the curvature (\ref{curvature}) for the new spin
connection one obtains
\begin{equation}
\mathbf{R}^{cd}(\mathbf{\omega}+\delta \mathbf{\omega}) \ = \
\mathbf{R}^{cd}(\mathbf{\omega )}\,+\,D\,\delta \mathbf{\omega
}^{cd}\,+\,\delta \mathbf{\omega }^{ce}\wedge \delta \mathbf{\omega
}_{e}^{\ d}\ ,\ \
\end{equation}
where $D$ is the covariant derivative defined by the original spin
connection. So long as we start from Minkowski space-time, then it
is $\mathbf{R}^{cd}(\mathbf{\omega )}=0$. Besides, it can be easily
verified
that the changes (\ref{change}), (\ref{deltaomega}) satisfy%
\begin{equation}
\epsilon _{abcd}\ \delta \mathbf{n}^{b}\wedge \delta
\mathbf{\omega}^{ce}\wedge \delta \mathbf{\omega }_{e}^{\ d}\ =\ 0\
. \label{cubic}
\end{equation}
In fact, it is
\begin{equation}
\delta \mathbf{\omega }^{ce}\wedge \delta \mathbf{\omega }_{e}^{\ d}\ \,=\
\left(
\begin{array}{rrrr}
0 & 0 & \ \ \mathbf{A} & \overline{\mathbf{A}} \\
0 & 0 & \ \ 0 & 0 \\
-\mathbf{A} & 0 & \ \ 0 & 0 \\
-\overline{\mathbf{A}} & 0 & \ \ 0 & 0%
\end{array}
\right) \ \ ,  \label{deltaomegadeltaomega}
\end{equation}
where
\begin{equation}
\mathbf{A}\ =\ \frac{f}{2\ \xi }\left( \partial _{\xi
}f+\partial_{\overline{\xi }}f\mathbf{-}\ \frac{f}{\xi
}+\frac{f}{\overline{\xi }}\right) \mathbf{n}\wedge \mathbf{m}\ .
\label{A}
\end{equation}%
So, according to Eq.~(\ref{deltaomegadeltaomega}), not only the
index $b$, but $c$ or $d$ should be zero to have a non-null
contribution to Eq.~(\ref{cubic}); since $b$, $c$, $d$ are
antisymmetrized, then Eq.~(\ref{cubic}) is satisfied.

Among the remaining terms in Eq.~(\ref{einsteinchange}), one of them
is linear in $f$ and the other ones are quadratic in $f$. They
should separately cancel out:
\begin{equation}
\epsilon _{abcd}\ \mathbf{n}^{b}\wedge D\,\delta \mathbf{\omega }^{cd}\ =\
0\ ,  \label{linear}
\end{equation}
\begin{equation}
\epsilon _{abcd}\ (\mathbf{n}^{b}\wedge \delta \mathbf{\omega }^{ce}\wedge
\delta \mathbf{\omega }_{e}^{\ d}+\delta \mathbf{n}^{b}\wedge D\,\delta
\mathbf{\omega }^{cd})\ =\ 0\ .  \label{quadratic}
\end{equation}
Let us begin by considering the contraction of these equations with
$\mathbf{n}^{a}$, which amounts the preservation of the value of the
Lagrangian (i.e., the preservation of the null scalar curvature).
Eq.~(\ref{quadratic}) is automatically accomplished when contracted
with $\mathbf{n}^{a}$. In fact, $\epsilon _{abcd}\
\mathbf{n}^{a}\wedge \mathbf{n}^{b}\wedge \delta \mathbf{\omega
}^{ce}\wedge \delta \mathbf{\omega }_{e}^{\ d}$ vanishes because
$(c,d)$ should be $(0,2)$ or $(0,3)$ (see
Eq.~(\ref{deltaomegadeltaomega})); so, $a$ or $b$ should be $1$,
what cancels out the expression (see Eq.~(\ref{A})). Besides,
$\epsilon _{abcd}\ \mathbf{n}^{a}\wedge \delta \mathbf{n}^{b}\wedge
D\,\delta \mathbf{\omega }^{cd}$ vanishes as a consequence of the
Eq.~(\ref{linear}). On the other hand, the contraction of
Eq.~(\ref{linear}) with $\mathbf{n}^{a}$ becomes
\begin{equation}
D(\epsilon _{abcd}\ \mathbf{n}^{a}\wedge \mathbf{n}^{b}\wedge \delta
\mathbf{\omega }^{cd})\ =\ 0\ ,  \label{derivada}
\end{equation}
where we have used that $D\,\mathbf{n}^{b}=0$ (null torsion) and
$D\epsilon _{abcd}=0$ (metricity). Notice that
\begin{equation}
\epsilon _{abcd}\ \mathbf{n}^{a}\wedge \mathbf{n}^{b}\wedge \delta
\mathbf{\omega }^{cd}\ =\ 2\sqrt{2}\left[ \left( \frac{1}{\xi
}+\frac{1}{\overline{\xi }}\right) \ f+\partial _{\xi }f+\partial
_{\overline{\xi }}f\right] \ \mathbf{n}\wedge \mathbf{m}\wedge
\mathbf{\overline{m}}\ ,
\end{equation}
which is a scalar-valued 3-form. So, no difference exists between $D$ and $d$
in Eq.~(\ref{derivada}), which reads
\begin{equation}
d\left[ \left( \frac{1}{\xi }+\frac{1}{\overline{\xi }}\right) \
f+\partial _{\xi }f+\partial _{\overline{\xi }}f\right] \wedge
\mathbf{n}\wedge \mathbf{m}\wedge \mathbf{\overline{m}\ +\ }\left[
\left( \frac{1}{\xi }+\frac{1}{\overline{\xi }}\right) \ f+\partial
_{\xi }f+\partial _{\overline{\xi }}f \right] \ d(\mathbf{n}\wedge
\mathbf{m}\wedge \mathbf{\overline{m}})\ =\ 0\ .
\end{equation}
We use Eqs.~(\ref{derivativesoftetrad}) and (\ref{dxi}) to obtain an
equation for $f$:
\begin{equation}
\frac{f}{\xi \ \overline{\xi }}+\left( \frac{1}{\xi
}+\frac{1}{\overline{\xi }}\right) \left( \partial _{\xi }+\partial
_{\overline{\xi }}\right) f+\frac{1}{2}\left( \partial _{\xi
}+\partial _{\overline{\xi }}\right) ^{2}f\ =\ 0\ .
\end{equation}
This equation just constrains $f$ to preserve the vanishing scalar
curvature. So, it is useful not only for vacuum Einstein equations
but also for traceless sources. The equation does not completely
determine $f$. Since just the operator $\partial _{\xi }+\partial
_{\overline{\xi }}=2$ $\partial _{\xi +\overline{\xi }}|_{_{\xi
-\overline{\xi }}}$ appears in this linear and homogeneous equation
for $f$, then the solution $f$ is undetermined by a factor $g(\xi
-\overline{\xi })$. The general solution of this equation is
\begin{equation}
f(\xi ,\overline{\xi })\ =\ \left[ \frac{Q^{2}}{\xi \ \bar{\xi}}-M\ \left(
\frac{1}{\xi }+\frac{1}{\bar{\xi}}\right) \right] \ g(\xi -\overline{\xi })\
,  \label{fuentesintraza}
\end{equation}
where $M$, $Q^{2}$ are integration constants. For a constant $g$, one would
obtain the Kerr-Newman metric,
\begin{eqnarray}
ds^{2} &=&dt^{2}-(dr\,+\,a\,\sin ^{2}\theta \ d\phi
)^{2}\,-\,(r^{2}+a^{2}\cos ^{2}\theta )\ (d\theta ^{2}\,+\,\sin
^{2}\theta \ d\phi ^{2})  \notag \\ \notag \\ &&-\left[ M\ \left(
\frac{1}{\xi }+\frac{1}{\bar{\xi}}\right) -\frac{Q^{2}}{\xi \
\bar{\xi}}\right] \ \left( dt-dr\,-\,a\,\sin ^{2}\theta \ d\phi
\,\right) ^{2}\ , \label{Kerr-Newman}
\end{eqnarray}
expressed in coordinates similar to the ones Kerr used in his
original article \cite{Kerr63,Kerr07} ($u\doteq t-r,\,r,\,\theta
,\,\phi $ are outgoing Eddington-Finkelstein-like coordinates;
ingoing coordinates would result if the starting point at
Eq.~(\ref{change}) were $\mathbf{n}\,\rightarrow \,\mathbf{n}\,\
+\,f\,\mathbf{l}$). Other coordinatizations can be found in
Ref.~\cite{Visser07}. In Eq.~(\ref{fuentesintraza}) one recognizes
the Newman-Janis rules to pass from Reissner-Nordstr\"{o}m to
Kerr-Newman:
\bigskip
\begin{eqnarray*}
&&\text{iii) in the gravitational potential, replace \ \ \ \ \ \ \ \
\ \ \ }
\frac{2}{r}\ \longrightarrow \ \frac{1}{\xi }+\frac{1}{\bar{\xi}}\,\,\,, \\
&&\text{iv) in the electric term, replace \ \ \ \ \ \ \ \ \ \ \ \ \
\ \ \ \ \ \ \ \ \ \ \ }\frac{1}{r^{2}}\ \longrightarrow \
\frac{1}{\xi \ \bar{\xi}} \,\,\,.
\end{eqnarray*}
It could be said that Newman-Janis algorithm does work because the
rest of Einstein equations constrain the function $g(\xi -
\overline{\xi })$ in Eq.~(\ref{fuentesintraza}) to be a constant.
Otherwise, function $f(\xi ,\bar{\xi})$ would contain a dependence
on $\xi -\overline{\xi }=2\,i\,a\,\cos \theta $ with no trace of it
in Schwarzschild or Reissner-Nordstrom solutions.

\subsection{More equations for $f(\protect\xi ,\bar{\protect\xi})$}
The story has not finished yet, because $f$ must still accomplish
the rest of the equations involved in Eqs.~(\ref{linear}) and
(\ref{quadratic}). Since we are trying with vacuum solutions, the
electric term in Eq.~(\ref{fuentesintraza}) should be suppressed by
Einstein equations. Besides, function $g(\xi -\overline{\xi})$
should be constrained to be a constant. One can easily verify that
three of the four equations in Eq.~(\ref{quadratic}) --those for
$a=0,2,3$-- are trivially
fulfilled by the matrices (\ref{omega}), (\ref{deltaomega}) and (\ref%
{deltaomegadeltaomega}). Since we have already worked a combination
of the equations taking part in Eq.~(\ref{quadratic}), it only
remains to satisfy Eq.~(\ref{linear}), which reads
\begin{equation}
DV_{a}=dV_{a}-\mathbf{\omega }_{\ a}^{b}\wedge V_{b}=0\
,\hspace{2cm}\text{where \ }V_{a}\doteq \epsilon _{abcd}\
\mathbf{n}^{b}\wedge \delta \mathbf{\omega}^{cd}\ .
\label{theequation}
\end{equation}%
According to the expression (\ref{deltaomega}) for $\delta \omega ^{ab}$,
the covector-valued 2-form $V_{a}$ is
\begin{equation}
V_{a}\ =\ 2\,\epsilon _{a\beta 0\gamma }\,\mathbf{n}^{\beta }\wedge \delta
\omega ^{0\gamma }\,+\,2\,\epsilon _{ab23}\,\mathbf{n}^{b}\wedge \delta
\omega ^{23}\,.
\end{equation}

\bigskip

i) $a=0$

\bigskip $V_{0}=0$. Thus the $0$-component of $DV_{a}=0$ is:
\begin{equation}
0\mathbf{\ }=\mathbf{\ \omega }_{\ 0}^{b}\wedge V_{b}\mathbf{\ }=\mathbf{\
\omega }^{21}\wedge V_{2}\mathbf{\ }+\mathbf{\ \omega }^{31}\wedge V_{3}\ ,
\label{V0}
\end{equation}
where $\mathbf{\omega }^{21}=\frac{\mathbf{m}}{\sqrt{2}\
\overline{\xi }}=\overline{\mathbf{\omega }^{31}}$, and
\begin{equation}
-V_{2}\,=\,\overline{V_{3}}\,=\,\sqrt{2}\left(
\frac{f}{\overline{\xi }}+\partial _{\xi }f+\partial _{\overline{\xi
}}f\right) \,\mathbf{n}\wedge \mathbf{\overline{m}}\ .
\end{equation}
Thus, $f(\xi ,\bar{\xi})$ must fulfill
\begin{equation}
\left( \frac{1}{\xi ^{2}}+\frac{1}{\overline{\xi }^{2}}\right) \ f\
+\ \left( \frac{1}{\xi }+\frac{1}{\overline{\xi }}\right) \left(
\partial _{\xi }+\partial _{\overline{\xi }}\right)f \ =\ 0\ , \label{fsincarga}
\end{equation}
whose general solution does not contain the electric charge term:
\begin{equation}
f\ =\ \left( \frac{1}{\xi }+\frac{1}{\overline{\xi }}\right) \
g(\xi-\overline{\xi })\ .  \label{resultado}
\end{equation}

\bigskip

ii) $a=1$

\smallskip

The $1$-component of $DV_{a}=0$ is:
\begin{equation}
0\ =\ dV_{1}\,-\,\mathbf{\omega }^{10}\wedge
V_{1}\,-\,\mathbf{\omega }^{20}\wedge
V_{2}\,-\,\mathbf{\omega}^{30}\wedge V_{3}\ =\ dV_{1}\,+\,
\mathbf{\omega }^{21}\wedge V_{2}\,+\,\mathbf{\omega }^{31}\wedge
V_{3} \label{V1}
\end{equation}
(see Eq.~(\ref{omega})). The last two terms cancel out because the
function $f$ verifies the Eq.~(\ref{V0}). So, the Eq.~(\ref{V1})
says that $V_{1}$ is a closed 2-form: $dV_{1}=0$, where
\begin{eqnarray}
V_{1}\ &=&\ 2\,\epsilon _{1203}\,\mathbf{m}\wedge \delta \omega
^{03}\,+\,2\,\epsilon _{1302}\,\overline{\mathbf{m}}\wedge \delta \omega
^{02}\,+\,2\,\epsilon _{1023}\,\mathbf{l}\wedge \delta \omega ^{23}  \notag
\\
\\
&=&2(\partial _{\xi }f\ d\xi -\partial _{\overline{\xi }}f\
d\overline{\xi})\wedge \mathbf{n\ }-\ \sqrt{2}\ \left( \frac{f}{\xi
}-\frac{f}{\overline{\xi}}+\partial _{\xi}f-\partial _{\overline{\xi
}}f\right) \ \mathbf{l} \wedge \mathbf{n}+\sqrt{2}f\ \left(
\frac{1}{\xi }+\frac{1}{\overline{\xi }} \right) \ \mathbf{m}\wedge
\overline{\mathbf{m}}  \notag
\end{eqnarray}
and $d\xi $ is given in Eq.~(\ref{dxi}). Let us examine the
component $\mathbf{l\wedge n}\wedge \mathbf{m}$ of the 3-form
$dV_{1}$. Noticeably, $d\mathbf{n}$, $d(\mathbf{l}\wedge
\mathbf{n)}$, $d(\mathbf{m}\wedge \overline{\mathbf{m}})$ and \
$d\xi \wedge d\overline{\xi }\wedge \mathbf{n}$ do not contribute to
such component. Instead,
\begin{equation}
d\overline{\xi }\wedge \mathbf{l}\wedge \mathbf{n}\ =\
\frac{i\sqrt{2} \,a\,\sin \theta }{\xi }\ \mathbf{l\wedge n}\wedge
\mathbf{m}\ .
\end{equation}
Therefore the component $\mathbf{l\wedge n}\wedge \mathbf{m}$ of $dV_{1}$
vanishes if%
\begin{equation}
\partial _{\overline{\xi }}\left[ \left( \frac{1}{\xi }-
\frac{1}{\overline{\xi}}\right) \ f+\partial _{\xi }f-
\partial _{\overline{\xi }}f\right] =0\ .
\label{resultado2}
\end{equation}
By replacing the result (\ref{resultado}) in Eq.~(\ref{resultado2})
one gets
\begin{equation}
\partial _{\overline{\xi }}\left[ 2\left( \frac{1}{\xi }+
\frac{1}{\overline{\xi }}\right) \ g^{\prime }(\xi -\overline{\xi })\right] =0\ .
\end{equation}
A similar complex conjugate equation is obtained by analyzing the
component $\mathbf{l\wedge n}\wedge \overline{\mathbf{m}}$ of
$dV_{1}$. Therefore, function $g$ is constant. The reader can verify
that $f(\xi ,\overline{\xi } )=-M\ (\frac{1}{\xi
}+\frac{1}{\bar{\xi}})$ also cancels the rest of the components of
$dV_{1}$ as well as those of $DV_{a}$ for $a=2,3$.

\section{Conclusion}\label{sec:V}
Reissner-Nordstrom geometry written in Kerr-Schild form,
\begin{equation}
\mathbf{g}\ =\ \mathbf{l}\otimes \mathbf{n}\,+\,\mathbf{n}\otimes
\mathbf{l}\,-\, \mathbf{m}\otimes
\overline{\mathbf{m}}\,-\,\overline{\mathbf{m}}\otimes \mathbf{m} \
-\,2\ \left( \frac{2M}{r}-\frac{Q^{2}}{r^{2}}\right) \,\mathbf{n}
\otimes \mathbf{n}\ ,  \label{SchKerrSchild}
\end{equation}%
where $\mathbf{l}\,=2^{-\frac{1}{2}}(dt + dr)$,
$\mathbf{n}\,=2^{-\frac{1}{2}}(dt - dr)$,
$\mathbf{m}\,=2^{-\frac{1}{2}}r(d\theta + i\,\sin \theta \ d\phi )$,
is promoted to Kerr-Newman geometry through the rules\bigskip
\begin{eqnarray*}
&&\text{i) in the }\mathbf{\{m,\overline{\mathbf{m}}\}}\text{
sector, replace \ \ \ \ \ \ \ \ \ }r\ \longrightarrow \ \xi\,,\
\overline{\xi}\ \ (\text{where}\ \ \xi \doteq r\,+\,i\ a\,\cos
\theta) \\
\\
&&\text{ii) in the }\mathbf{\{l,n\}}\text{ sector, replace \ \ \ \ \ \ \ \ \
\ }dr\ \longrightarrow \ dr\,+\,a\,\sin ^{2}\theta \ d\phi \ \,, \\
\\
&&\text{iii, iv) replace \ \ \ \ \ \ \ \ \ \ \ \ \ \ \ \ \ \ \ \ \ \
\ \ \ \ \ \ \ \ \ \ } \frac{2M}{r}-\frac{Q^{2}}{r^{2}}\text{\ \
}\longrightarrow \text{\ \ \ }M\, \left( \frac{1}{\xi
}+\frac{1}{\bar{\xi}}\right) -\frac{Q^{2}}{\xi \ \bar{\xi}}\,\,\,\,.
\end{eqnarray*}
\medskip

Newman-Janis algorithm results to be a simple rule because Einstein
equations constrain function $g(\xi-\overline{\xi })$ in
Eqs.~(\ref{fuentesintraza}) and (\ref{resultado}) to be a constant.
Thus, any possible dependence on $\xi-\overline{\xi
}=2i\,a\,\cos\theta$ is excluded from Kerr-Newman geometry;
otherwise, Kerr-Newman geometry would contain a dependence on a
variable leaving no trace in the $a=0$ Schwarzschild geometry. It
can be concluded that Newman-Janis algorithm seems to be linked to
particular features of Einstein's theory that could hardly be
replicated in other theories.

\bigskip

As a final remark, notice that the $\mathbf{n}\otimes \mathbf{n}$
term in Eq.~(\ref{SchKerrSchild}) displays non-diagonal components
$g_{rt}$. The usual diagonal form of Reissner-Nordstrom metric
tensor is obtained by means of a redefinition of $t$. Likewise,
Kerr-Newman metric in Boyer-Lindquist coordinates is reached not
only by undoing the twisting (\ref{twisting}) but
by redefining the time too:%
\begin{equation}
d\varphi \ \doteq \ d\phi \,\,+\,\frac{a}{r^{2}+a^{2}-2Mr+Q^{2}}\ dr\ ,
\label{varphi}
\end{equation}
\begin{equation}
d\widetilde{t}\ \doteq \ dt\,\,+\,\frac{2Mr-Q^{2}}{r^{2}+a^{2}-2Mr+Q^{2}}\
dr\ .  \label{time}
\end{equation}
Thus, the Kerr-Newman geometry (\ref{Kerr-Newman}) in Boyer-Lindquist
coordinates is

\begin{equation}
ds^{2}\ =\ d\widetilde{t}\ ^{2}-\frac{2Mr-Q^{2}}{\rho ^{2}}\ \left(
d \widetilde{t}-a\,\sin ^{2}\theta \ d\varphi \right)
^{2}-\frac{dr^{2}}{\frac{r^{2}+a^{2}}{\rho
^{2}}-\frac{2Mr-Q^{2}}{\rho ^{2}}}\ -\,\rho ^{2}\ d\theta ^{2}\,-\
\left( r^{2}+a^{2}\right) \ \sin ^{2}\theta \ d\varphi ^{2}\ ,
\end{equation}
\vskip.2cm\parindent0pt where $\rho^2\,\doteq
\,r^2+a^2\cos^2\theta$. For $a=0$, the usual form of
Reissner-Nordstrom geometry is recovered.



\end{document}